\documentclass[aps,prb,twocolumn,superscriptaddress]{revtex4-2}
\usepackage{amsmath,amsfonts,amssymb}
\usepackage{graphicx,float,calc}
\usepackage{color,bm}
\usepackage{ulem}
\usepackage{braket}
\usepackage[colorlinks,urlcolor=blue,citecolor=blue,linkcolor=blue]{hyperref}

\begin{document}

\title{Stabilizing confined quasiparticle dynamics in one-dimensional polar lattice gases}

\author{Guo-Qing Zhang}
\email{zhangptnoone@zjhu.edu.cn}
\affiliation{Research Center for Quantum Physics, Huzhou University, Huzhou 313000, P. R. China}
\author{L. F. Quezada}
\email{lf\_quezada@outlook.com}
\affiliation{Research Center for Quantum Physics, Huzhou University, Huzhou 313000, P. R. China}
\affiliation{Laboratorio de Ciencias de la Informaci\'{o}n Cu\'{a}ntica, Centro de Investigaci\'{o}n en Computaci\'{o}n, Instituto Polit\'{e}cnico Nacional, UPALM, 07700, Ciudad de M\'exico, M\'exico}

\date{\today}

\begin{abstract}
The disorder-free localization that occurred in the study of relaxation dynamics in far-from-equilibrium quantum systems has been widely explored. Here we investigate the interplay between the dipole-dipole interaction (DDI) and disorder in the hard-core polar bosons in a one-dimensional lattice. We find that the localized dynamics will eventually thermalize in the clean gas, but can be stabilized with the existence of a small disorder proportional to the inverse of DDI strength. From the effective dimer Hamiltonian, we show that the effective second-order hopping of quasiparticles between nearest-neighbor sites is suppressed by the disorder with strength similar to the effective hopping amplitude. The significant gap between the largest two eigenvalues of the entanglement spectrum indicates the dynamical confinement. We also find that the disorder related sample-to-sample fluctuation is suppressed by the DDI. Finally, we extend our research from the uncorrelated random disorder to the correlated quasiperiodic disorder {\color{black}and from the two-dimer model to the half-filling system, obtaining} similar results.
\end{abstract}

\maketitle

\section{Introduction}
Recently,  exotic phenomena from the studies of relaxation dynamics in far-from-equilibrium quantum systems have been unraveled. The dynamical confinement has been shown to exist in the quantum quench dynamics of lattice models with the $1/r^3$ dipolar tail interaction~\cite{PhysRevB.92.180406,PhysRevLett.124.010404,PhysRevLett.126.023001,PhysRevLett.127.260601,PhysRevA.107.013301} and short-range interacting spin chains with both transverse and longitudinal fields~\cite{kormos2017real,PhysRevLett.122.130603,PhysRevLett.122.150601,PhysRevB.99.180302}. The disorder-free localization (DFL) emergent from dynamical confinement extends the phenomenology of disorder induced many-body localization (MBL)~\cite{PhysRevLett.44.1288,PhysRevLett.73.2607,PhysRevB.82.174411}.  A finite number of conservation laws lead to the fragmentation of Hilbert space in DFL which severely constrains the dynamics and violates the eigenstate thermalization hypothesis (ETH)~\cite{10.1080/00018732.2016.1198134}, while the emergent of local integrals of motion is the ingredient of no thermalization in disorder MBL~\cite{PhysRevA.74.053616,PhysRevLett.98.050405}.  MBL and DFL provide different mechanisms to violate ETH and localize the quench dynamics of local observables.

The interest in quench dynamics has been roused by breakthrough experimental developments in recent years.  Isolated many-body quantum systems can be almost perfectly realized in cold gases and trapped ions~\cite{kinoshita2006quantum,gring2012relaxation,richerme2014non,garttner2017measuring} with naturally occurring long-range interactions.  These experiments unveil the possibility of simulating a wide variety of quantum many-body quench dynamics.  The extended Hubbard model with nearest-neighbor interactions has already been realized on polar lattice gas experiments~\cite{yan2013observation,PhysRevLett.111.185305}.  For the paradigmatic interactions decaying as a power law $1/r^3$ with distance $r$, this term can be generalized in polar gases with strong dipole-dipole interactions (DDIs)~\cite{ni2008high,ni2010dipolar,PhysRevLett.108.080405}, or Rydberg atoms with strong van der Waals interactions~\cite{RevModPhys.82.2313,schauss2012observation,PhysRevLett.110.263201}.  Such long-range interactions may host novel properties which are missing in their short-ranged counterparts~\cite{PhysRevLett.119.023001}.

For the quench dynamics of DFL systems, a small effective second-order hopping or a certain degree of gauge-breaking errors can create transitions between isolated sectors in the Hilbert space and finally undermine the quasiparticle confinement\cite{PRXQuantum.3.020345,halimeh2021stabilizing}.  At the long-time limit, the quasilocalized state will eventually thermalize regardless of the error strength.  It is interesting to investigate the stabilizing of the dynamical confinement in a polar bosons lattice model with DDI, and reveal the interplay between long-range interactions and quenched disorders.  For a one-dimensional (1D) system, the low-density filling of the extended Hubbard model corresponds to strong correlations, which present rich physics in the few-body problem~\cite{PhysRevLett.126.023001}, so we investigate the dynamics of two dimers (quasiparticles) in the extended boson-Hubbard model.  In this paper, we adopt half-chain entanglement, out-of-time-ordered correlation (OTOC), inhomogeneity parameter, and return probability to characterize the dynamical confinement.  All four quantities evidence the thermalization of DFL in the long-time limit for a clean polar gas. However, it is revealed that a very small disorder can stabilize DFL in the long-time dynamics, where the disorder induced localization is ignorable. We use a dimer approximation to unveil the stabilization. The disorder strength needed to suppress the effective second-order hopping of dimers between nearest-neighbor sites is proportional to the inverse of the DDI strength. The entanglement spectra of long-time dynamical confined states indicate a significant gap between the largest and the second largest eigenvalues. We also find that the DDI can reduce sample-to-sample fluctuations which are usually logarithmically broad in the disorder induced localization. Finally, we extend our results from the uncorrelated random disorder to the infinitely correlated quasiperiodic disorder saturation, {\color{black} and to the half-filling case with more quasiparticles dynamically stabilized in the presence of a very weak disorder.}

The rest of this paper is organized as follows.  In Sec.~\ref{sec2}, we introduce the extended boson-Hubbard model with DDI and its effective approximation. Section \ref{sec3} is devoted to revealing the dynamical confinement stabilized via the existence of random disorder by four different physical quantities. We analyze the interplay of DDI and random disorder in Sec.~\ref{sec4}, and unveil the same result for the quasiperiodic disorder case in Sec.~\ref{sec5}. A brief discussion and conclusion are finally given in Sec.~\ref{secf}.

\section{\label{sec2}Model}
We consider a hard-core polar boson gas in a 1D optical lattice with long-range DDIs and disordered on-site potentials. The system is described by the extended boson-Hubbard Hamiltonian~\cite{PhysRevLett.127.260601}:
\begin{equation}\label{ebhm}
H=-J\sum_i(b_i^\dagger b_{i+1} + \mathrm{H.c.})+\sum_i \epsilon_i n_i + V \sum_{i<j}\frac{1}{|i-j|^3} n_i n_j,
\end{equation}
where $b_i$ ($b_i^\dagger$) is the annihilation (creation) operator of boson on site $i$ with hard-core condition $(b_i^\dagger)^2=0$, $n_i=b_i^\dagger b_i$ is the number operator, $J$ is the hopping amplitude, $\epsilon_i \in [-W, W]$ is the uniformly distributed random potential with disorder strength $W$, and $V$ is the DDI strength between nearest neighbors. We use exact diagonalization calculation to investigate the long-time quench dynamics governed by Eq.~(\ref{ebhm}) for a four-particle problem.

For moderate DDI strength $V$, two particles locate at two nearest sites can form a dynamically bound nearest-neighbor dimer (NND), and the dynamics are dominated by the effective  Hamiltonian defined in the dimer subspace~\cite{PhysRevLett.124.010404}.  For sufficiently large $V$, all particles are paired in NNDs, and we can write Eq.~(\ref{ebhm}) into NND bases which reads
\begin{equation}
\begin{split}\label{debhm}
H_d=&-J_d\sum_l(D_i^\dagger D_{i+1} + \mathrm{H.c.}) +\sum_i \epsilon'_i N_i \\
& + V \sum_{i,l>0}f(l)N_iN_{i+l+2},
\end{split}
\end{equation}
where $D_i^\dagger = b_i^\dagger b_{i+1}^\dagger$ is the creation operator of an NND at site $i$, $N_i=D^\dagger_i D_i$ is the number operator of NNDs, $J_d=8J^2/7V$ is the effective second-order hopping amplitude,  $\epsilon'_i=\epsilon_i+\epsilon_{i+1}$,  and $f(l)=2/(l+2)^3+1/(l+1)^3+1/(l+3)^3$ obtained from the DDI between two dimers separated by distance $l$. The quasiparticle can move from site $i$ to $i+1$ in a second-order procedure. Let us consider a configuration $110$ where $1$ stands for occupied boson and $0$ for empty site.  The NND can move forward to the right-hand site in the following procedure $110\rightarrow 101 \rightarrow 011$ with amplitude $J_d=J^2/(V-V/2^3)=8J^2/7V$~\cite{PhysRevB.92.180406}. The initial four-boson evolution under Eq.~(\ref{ebhm}) can be reduced to a two-dimer dynamics described by the effective Hamiltonian~(\ref{debhm}).

\section{\label{sec3}Dynamical confinements}
\begin{figure}[t!]
\centering
\includegraphics[width=0.48\textwidth]{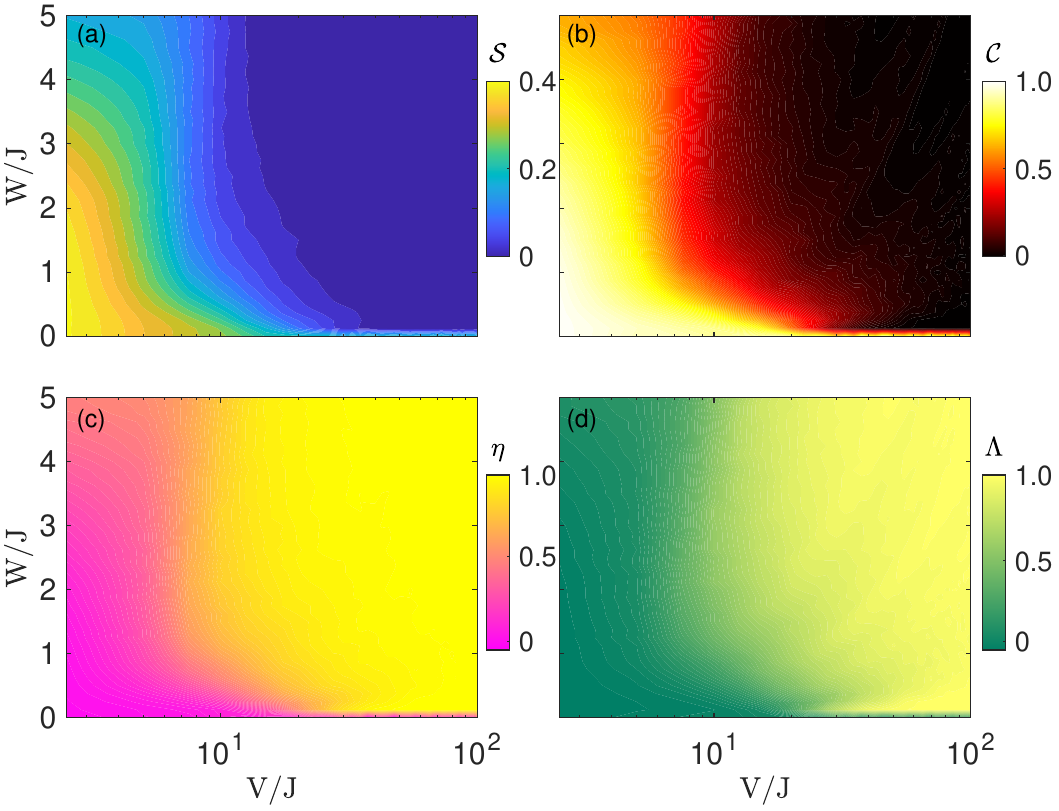}
\caption{(Color online) Long-time saturated values of (a) half-chain entanglement $\mathcal{S}$ , (b) OTOC $\mathcal{C}$, (c) inhomogeneity parameter $\eta$, and (d) return probability $\Lambda$ plotted in the $V$-$W$ plane. All data are averaged from time interval $t\in [10^6, 10^8]$ and 100 disorder realizations. The chain length is $L=16$.}
\label{figall}
\end{figure}

We first consider the long-time dynamics of the Hamiltonian (\ref{ebhm}) with both DDI and disorder. In the clean limit ($W=0$) and sufficiently large $V$, two dimers initially located within a critical distance will stay fixed for a certain time and eventually thermalize due to the effective second-order hopping~\cite{PhysRevLett.124.010404}. In the following, we consider the system initially prepared as two dimers separated by four sites and exactly calculate the time dependent wave function using Eq.~(\ref{ebhm}). The initial state reads $\ket{\psi_0}=\ket{\cdots 0 11 0000 11 0 \cdots}$ in the Fock state basis, and the time-dependent wave function is $\ket{\psi(t)}=\mathrm{exp}(-iHt)\ket{\psi_0}$. We adopt four different quantities,  half-chain entanglement $\mathcal{S}$, OTOC $\mathcal{C}$, inhomogeneity parameter $\eta$, and return probability $\Lambda$ to consistently characterize the localization and dynamical confinement. The open-boundary condition is assumed in our calculation.

\subsection{Half-chain entanglement}
\begin{figure*}[t!]
\centering
\includegraphics[width=0.8\textwidth]{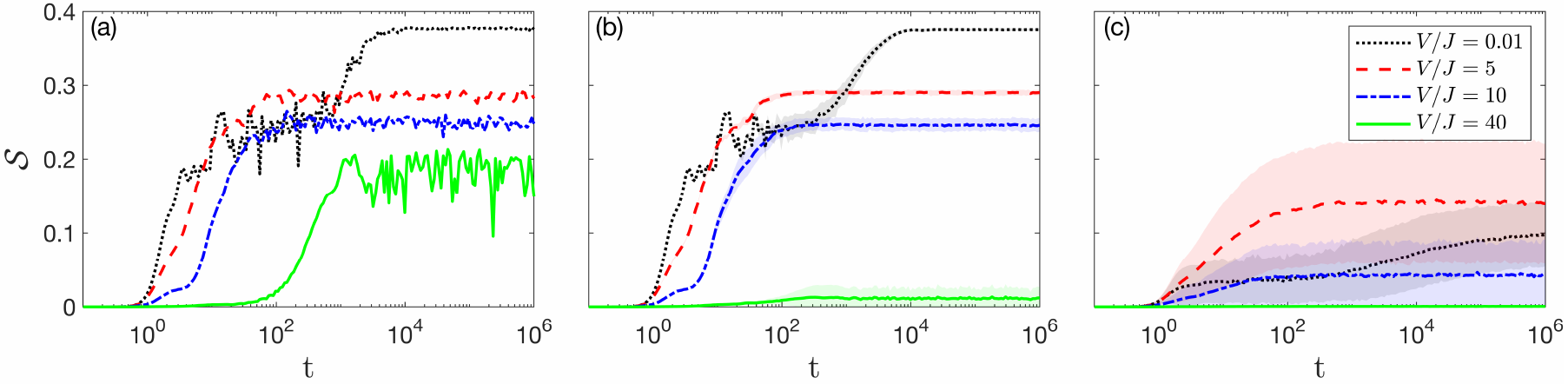}
\caption{(Color online) Long-time growth of half-chain entanglement $\mathcal{S}$ displayed as a function of evolution time $t$. $V=0.01$ for black dotted lines, $V=5$ for red dashed lines, $V=10$ for blue dot-dashed lines, and $V=40$ for green solid lines. Disorder strengths are (a) $W=0$, (b) $W=0.1$, and (c) $W=5$. Faded colors indicate the standard error between different samples. The chain length is $L=20$, and 100 disorder realizations are averaged for $W\neq0$ curves.}
\label{figee}
\end{figure*}

The half-chain entanglement is a commonly used entropy whose slow-down dynamics and saturated value can reveal localization properties of the system~\cite{PhysRevLett.109.017202,PhysRevResearch.2.032039,10.1080/00018732.2016.1198134,PhysRevB.102.054204}. We can express quantum states under the bases of two subsystems $\ket{\psi(t)}=\sum_{ij}\delta_{ij}(t)\ket{\psi^A_i}\otimes\ket{\psi^B_j}$, where $\ket{\psi^A_i}$ ($\ket{\psi^B_j}$) is the subspace basis of the left (right) half chain. The reduced density matrix element of the left subsystem is then obtained as $\rho^A_{ii'}(t)=\sum_j \delta_{ij}(t) \delta^*_{i'j}(t)$, and the half-chain entanglement per site is defined as the von Neumann entropy of the reduced density matrix
\begin{equation}
\mathcal{S}(t)=-\frac{1}{L_A}\mathrm{Tr}\rho^A(t) \ln(\rho^A(t)).
\end{equation}
One can use the reduced density matrix of the right half chain to obtain the same entanglement because $\rho^B$ shares the same spectrum as $\rho^A$.

We consider a chain length $L=16$ with a bipartition of equal half $L_A=L_B=L/2$ to calculate $\mathcal{S}$. Hamiltonian~(\ref{ebhm}) is exactly diagonalized to obtain the long-time many-body wave function $\ket{\psi(t)}$. In Fig.~\ref{figall} (a), we present the saturated value of $\mathcal{S}$ after the quench dynamics which is averaged from time interval $t\in[10^6, 10^8]$ as a function of DDI $V$ and disorder $W$. For $W\neq 0$, 100 random disorder configurations are used. The initial state $\ket{\psi_0}$ is a product state with vanishing entanglement, and for a well confined dynamics, the entanglement should be close to zero. We can observe in Fig.~\ref{figall} (a) that when $W=0$, the saturated value of $\mathcal{S}$ is significantly different from zero for any $V$. Only in those regions with moderate $V$ and nonvanishing $W$, $\mathcal{S}$ remains close to zero. We further display the growth of half-chain entanglement $\mathcal{S}$ for $L=20$ in Fig.~\ref{figee} with several typical values of $V$ and $W$. The smallest $V=0.01$ rather than $0$ is due to the fact that noninteracting systems can not produce any entanglement for the product state $\ket{\psi_0}$. For the clean limit $W=0$ in Fig.~\ref{figee} (a), $\mathcal{S}$ tends to finite value for all values of DDI $V$. {\color{black}For $V=40$, $\mathcal{S}$ remains close to zero for a certain time because the dimer-dimer cluster is dynamically confined. When the system begins to thermalize due to the center-of-mass motion of the dimer-dimer cluster at $t\approx 10^2$, the entanglement grows steeply. The dimers begin to move at $t\sim 1/J_D\sim V$, and the rapid growth of the entanglement starts earlier for smaller values of $V$.} For a weak disorder $W=0.1$ in Fig.~\ref{figee} (b), the saturated value of entanglement for $V=40$ decreases significantly and the system shows a well confined dynamics, {\color{black}where a fixed distance between dimers is rigidly formed under the interplay of dipolar interaction and weak disorder.}  The saturated values for other values of $V$ are similar to the clean limit case, {\color{black}and the timescale for the rapid growth is not changed. In these cases, the particles can move as freely as the clean ones.} When increasing disorder $W$ to $5$ in Fig.~\ref{figee} (c), the disorder-induced localization comes into play for all values of $V$. Especially when $V=40$, $\mathcal{S}\approx 0$ in the entire time interval which reveals the perfect dynamical confinement.

\subsection{Out-of-time-ordered correlation}
\begin{figure*}[t!]
\centering
\includegraphics[width=0.8\textwidth]{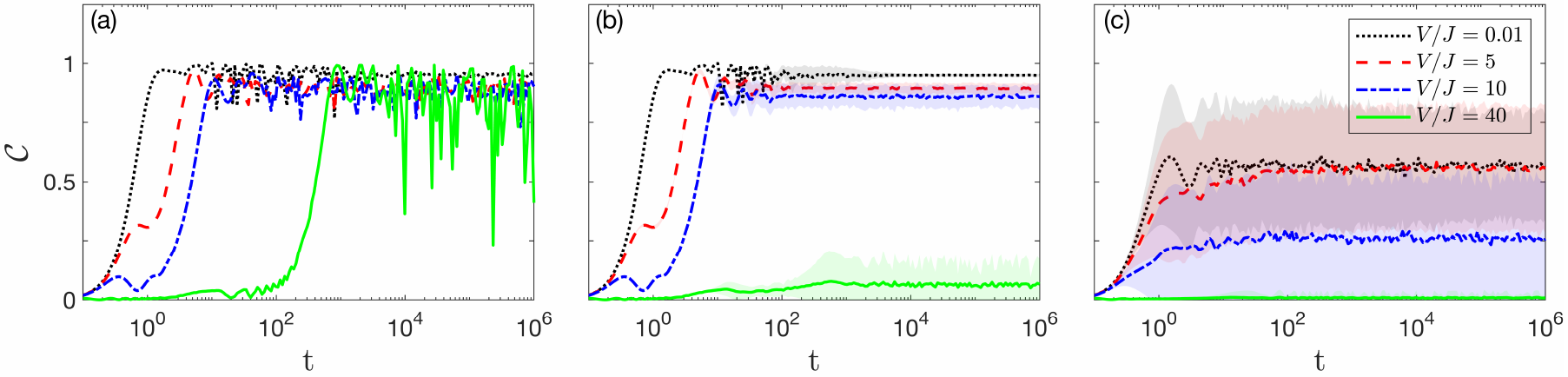}
\caption{(Color online) Long-time development of OTOC $\mathcal{C}$ presented as a function of time $t$. $V=0.01$ for black dotted lines, $V=5$ for red dashed lines, $V=10$ for blue dot-dashed lines, and $V=40$ for green solid lines. Disorder strength are (a) $W=0$, (b) $W=0.1$, and (c) $W=5$. Faded colors are the standard error between different disorder samples. The chain length is $L=20$, and 100 disorder realizations are averaged for $W\neq0$ data.}
\label{figotoc}
\end{figure*}

OTOC characterizes the delocalization or scrambling of quantum information, whereby it describes the process of an initially localized state spreading over all degrees of freedom in a quantum many-body system~\cite{hashimoto2017out,PhysRevB.95.054201,PhysRevLett.120.040402,PhysRevResearch.1.032039}. OTOC has been related to entanglement, and can serve as an experimentally accessible entanglement witness~\cite{hosur2016chaos,FAN2017707}. OTOC can also be used as an order parameter to characterize the localization-delocalization transition in disordered many-body systems~\cite{PhysRevB.95.054201,PhysRevB.99.184202,PhysRevB.105.205133}. The OTOC arises from the squared commutator of two commuting local operators $\hat{V}$ and $\hat{W}$:
\begin{equation}
\begin{split}
\mathcal{C}(t)=&\frac{1}{2}\braket{[\hat{V}(t),\hat{W}]^\dagger[\hat{V}(t),\hat{W}]}\\
=&\frac{1}{2}[\braket{\hat{W}^\dagger\hat{V}^\dagger(t)\hat{V}(t)\hat{W}}+\braket{\hat{V}^\dagger(t)\hat{W}^\dagger\hat{W}\hat{V}(t)}\\
 &-\braket{\hat{V}^\dagger(t)\hat{W}^\dagger\hat{V}(t)\hat{W}}-\braket{\hat{W}^\dagger\hat{V}^\dagger(t)\hat{W}\hat{V}(t)}].
\end{split}
\end{equation}
For Hermitian and unitary local operators, the OTOC can be simplified as~\cite{PhysRevB.97.144304}
\begin{equation}
\mathcal{C}(t)=1- \mathrm{Re}\braket{\hat{V}^\dagger(t)\hat{W}^\dagger\hat{V}(t)\hat{W}}.
\end{equation}
For confined dynamics, information spreading is suppressed and those two local operators commute at different times, which leads to the vanishment of OTOCs.

We choose two local operators as NND number operators of the initial state $\ket{\psi_0}$, $\hat{V}=n_in_{i+1}$ and $\hat{W}=n_{i+l+1}n_{i+l+2}$ in our following study. In Fig.~\ref{figall} (b), we plot the saturated value of OTOC $\mathcal{C}$ as a function of DDI $V$ and disorder $W$. Similar to the half-chain entanglement, OTOC remains close to zero in regions where $V$ is moderate and $W$ is not zero. We also show the long-time dynamics of $\mathcal{C}$ for $L=20$ in Fig.~\ref{figotoc} with all parameters the same as in Fig.~\ref{figee}. The clean limit in Fig.~\ref{figotoc} (a) indicates the system finally thermalizes for a long enough time $t$. When $V=40$ (green solid lines), the dynamics show well and perfect confinement for very small disorder $W=0.1$ [Fig.~\ref{figotoc}(b)] and large disorder $W=5$ [Fig.~\ref{figotoc}(c)], respectively. The localization properties induced by disorders appear for small $V$s when $W=5$, but the localization is not evident when disorder strength $W=0.1$. {\color{black}OTOC characterizes the propagation of localized information and can serve as measurable entanglement witness~\cite{hosur2016chaos,FAN2017707}. The behaviors of OTOC thus can interpreted in the same way as entanglement.}

\subsection{Inhomogeneity parameter}
\begin{figure*}[t!]
\centering
\includegraphics[width=0.8\textwidth]{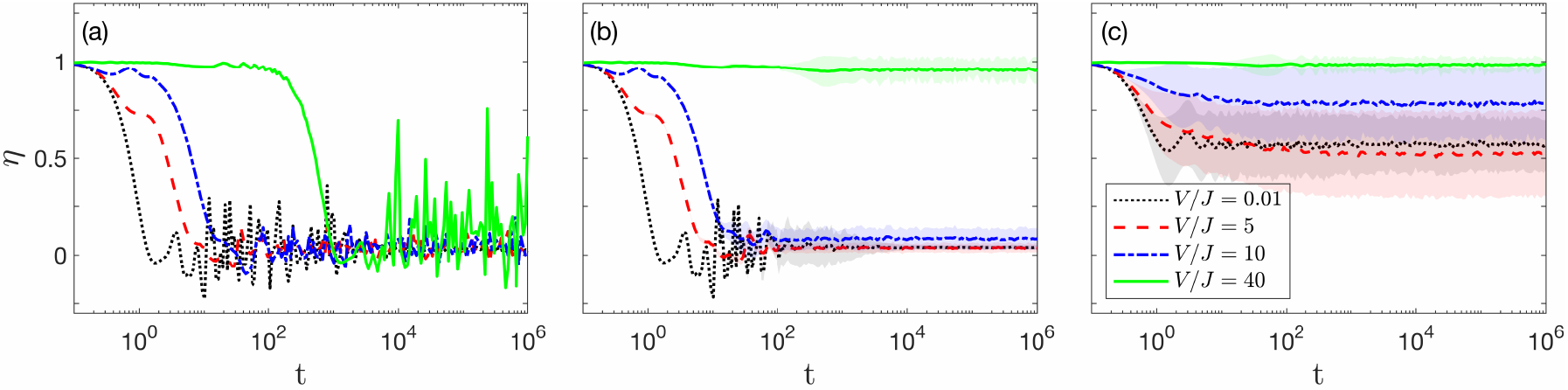}
\caption{(Color online) Long-time dynamics of inhomogeneity parameter $\eta$. $V=0.01$ for black dotted lines, $V=5$ for red dashed lines, $V=10$ for blue dot-dashed lines, and $V=40$ for green solid lines. Disorder strength are (a) $W=0$, (b) $W=0.1$, and (c) $W=5$. Faded colors represent the standard error between different disorder configurations. The chain length is $L=20$, and 100 disorder realizations are averaged for $W\neq0$ curves.}
\label{figeta}
\end{figure*}

The inhomogeneity parameter, similar to the imbalance for half-filling systems, characterizes the localization of particles~\cite{PhysRevLett.124.010404,PhysRevLett.127.260601}. The inhomogeneity parameter $\eta$ is defined in such a way that $\eta=1$ for the initial state and $\eta=0$ when density waves are uniformly distributed in the whole system:
\begin{equation}
\eta(t)=\frac{N_0(t)L_0^{-1}-N_bL^{-1}}{1-N_bL^{-1}},
\end{equation}
where $N_b=4$ is the total boson number, $L_0=4$ is the length of the occupied sites in the initial state, and $N_0(t)$ is the total particle number on the initially occupied sites after evolution time $t$.

We plot the saturated value of the inhomogeneity parameter $\eta$ averaged over the long-time dynamics interval $t\in[10^6, 10^8]$ in the $V$-$W$ plane in Fig.~\ref{figall} (c). When both $V$ and $W$ are small, $\eta$ is close to zero, which means the system becomes homogeneous and thermalized after long-time dynamics. For moderate $V$ but vanishing $W$, this system is delocalized due to weak second-order coupling between Fock states. When $W>0$, this region shows localization with inhomogeneity parameter $\eta\approx 1$. In Fig.~\ref{figeta}, the decreasing of $\eta$ is presented for $L=20$ systems with several typical values of interaction and disorder strength. Regarding the clean gas in Fig.~\ref{figeta} (a), the system can stay for longer periods of time in inhomogeneous with larger DDI $V$ before delocalization. {\color{black}$\eta$ is a direct indicator for the localization of particles, and the localization time of the rigidly formed dimer clusters can be directly seen from the plateau of $\eta$ before a steep decrease.} For a very small disorder $W=0.1$ in Fig.~\ref{figeta} (b), the green solid curve indicates inhomogeneity of the long-time dynamics and a long-lived memory of the initial condition for large $V=40$. {\color{black}The interplay of dipolar interaction and weak disorder prevent the center-of-mass motion of the dimer cluster, and no significant decrease occurs. Plateaus of the other three curves still dwindle quickly similar to the clean ones, where the disorder induced localization is not evident and the disorder effect can not prevent particles from moving.} For $W=5$ in Fig.~\ref{figeta} (c), disorder-induced localization takes into place and the saturated values of $\eta$ for $V=0.01, 5, 10$ improve significantly compared with those values in Fig.~\ref{figeta} (b). The interplay of DDI $V$ and disorder $W$ leads to inhomogeneity and thus makes dynamical confinement perfect.

\subsection{Return probability}
\begin{figure*}[t!]
\centering
\includegraphics[width=0.8\textwidth]{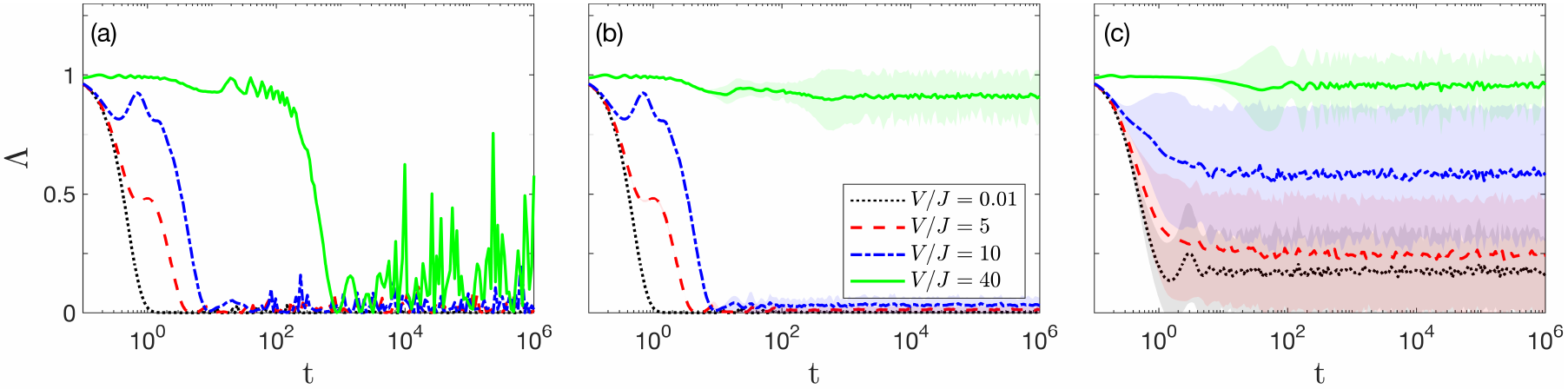}
\caption{(Color online) Long-time evolution of return probability $\Lambda$ as a function of time $t$. $V=0.01$ for black dotted lines, $V=5$ for red dashed lines, $V=10$ for blue dot-dashed lines, and $V=40$ for green solid lines. Disorder strength are (a) $W=0$, (b) $W=0.1$, and (c) $W=5$. Faded colors are the standard error. The chain length is $L=20$, and 100 disorder realizations are averaged for $W\neq0$ lines.}
\label{figrp}
\end{figure*}

Return probability determines the global property of time evolution which is widely used in the investigation of quench dynamics~\cite{PhysRevLett.96.140604,IZRAILEV2006355,PhysRevLett.110.135704,Stephan_2017,PhysRevB.98.134205}. Return probability is defined as the modulus squared of the overlap between the initial state $\ket{\psi_0}$ and the evolution wavefunction $\ket{\psi(t)}$ at time $t$:
\begin{equation}
\Lambda(t)=|\braket{\psi_0|\psi(t)}|^2=|\braket{\psi_0|e^{-iHt}|\psi_0}|^2.
\end{equation}
In the quench dynamics of thermalized systems, when the initial state is not close to any eigenstate of the Hamiltonian, this quantity is expected to tend to zero quickly in an exponential form $e^{-Lf(t)}$ with $L$ the system's size. $\Lambda$ can be considered as the probability to find the evolved system staying in the initial state after time $t$, and can be used to characterize localization and confined dynamics.

We show the saturated value of the return probability $\Lambda$ averaged over long-time interval as a function of $V$ and $W$ in Fig.~\ref{figall} (d). Similar to the previously studied three quantities, $\Lambda$ indicates that there is a small probability to find the initial state after a long-time evolution when both $V$ and $W$ are small, while the dynamics localized for moderate $V$ and nonzero $W$. The long-time dynamics of the return probability $\Lambda$ is plotted in Fig.~\ref{figrp} for $L=20$ sites with other parameters the same as in Fig.~\ref{figee}. Green solid curves in Figs.~\ref{figrp} (a) and (b) both show a certain time of confined dynamics when $t<10^2$ with $\Lambda$ close to one, but for $W=0$ the return probability suffers a steep reduction {\color{black}when center-of-mass motion plays the role at $t\sim 1/J_D$. The slow drop of $\Lambda$ before $t\sim 1/J_D$ lies in the fact that dimers are not bound rigidly to the original place but have a small probability to spread to their nearest neighbors.} The localization property induced by disorder is ignorable for $W=0.1$ in Fig.~\ref{figrp} (b) but noticeable in Fig.~\ref{figrp} (c) when $V=0.01, 5, 10$. Meanwhile, we can also observe the disorder-stabilized high return probability for $V=40$ in the long-time limit.

\section{\label{sec4}Interplay between dipole-dipole interaction and disorder}
In this section, we further investigate in detail the interplay between DDI and disorder in the extended Boson-Hubbard model. In the previous section, we have already revealed the long-time stabilized dynamical confinements with a very small disorder. This phenomenon can be interpreted by the approximated dimer Hamiltonian where all particles are paired in NNDs. We also discuss the entanglement spectrum of confined and unconfined wave functions whose properties are distinct. The sample-to-sample fluctuations for various disorders and interactions are revealed at the end of this section.

\subsection{Dynamically bound dimers under weak disorder}
\begin{figure}[t!]
\centering
\includegraphics[width=0.48\textwidth]{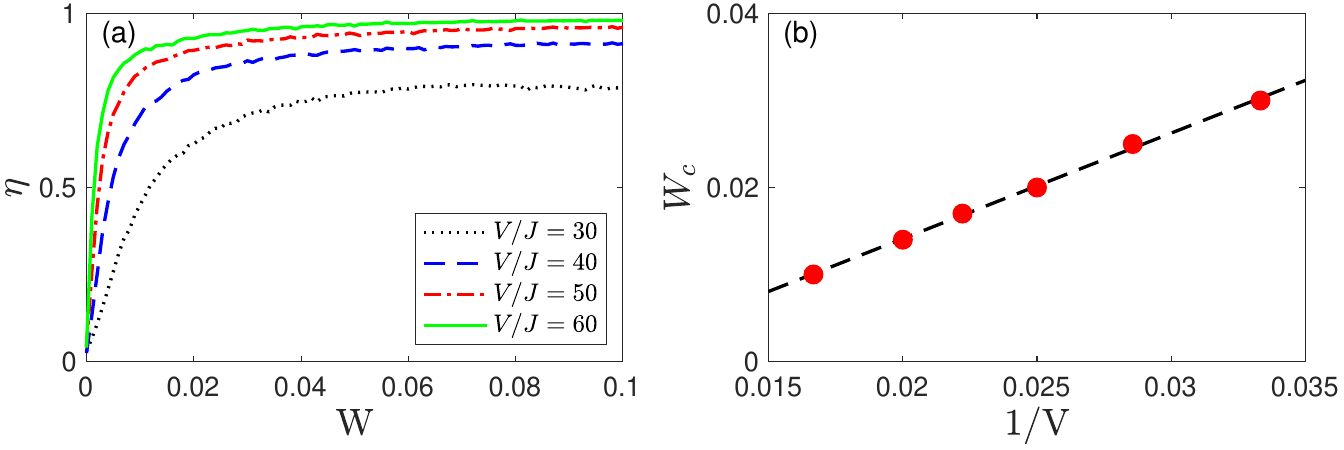}
\caption{(Color online) (a) Long-time saturated values of inhomogeneity parameter $\eta$ plotted as functions of disorder $W$. (b) The scaling of turning disorder strength $W_c$ as a function of $1/V$. Data are averaged from time interval $t\in [10^6, 10^8]$ and 100 disorder realizations. Dynamics are exactly calculated using Hamiltonian (\ref{debhm}), and NND chain length is $L=35$.}
\label{figetaw}
\end{figure}

For moderate to strong DDI, the dynamics of the initial state $\ket{\psi_0}$ under the extended boson-Hubbard model (\ref{ebhm}) can be approximated by the dimer extended boson-Hubbard Hamiltonian (\ref{debhm}) where two neighboring particles pair to an NND, and the quasiparticle NND can only move to a neighbor site via a second-order hopping process with amplitude $J_d=8J^2/7V$. In the clean limit, due to this small second-order hopping, the confined NNDs will eventually begin to spread over the whole system and thermalize after long-time dynamics. The initial state $\ket{\psi_0}$ is effectively two NNDs separated by five sites under the dimer bases. We exactly evaluate this initial state using Hamiltonian (\ref{debhm}), and investigate the saturated value of the inhomogeneity parameter $\eta$ as functions of disorder $W$. For small disorder varying from $0$ to $0.1$, we display the saturation $\eta$ for $V$ varying from $30$ to $60$ in Fig.~\ref{figetaw} (a). As we can see, $\eta$ increases rapidly and saturates for a very small disorder $W$ which is comparable to the second-order hopping amplitude $J_d\sim 1/V$. Due to disorder fluctuations, $\eta$ curves are not smooth enough to do numerical derivatives with respect to $W$. Thus, we define the turning disorder strength $W_c$ as the value where $\eta(W_c)=0.9\eta(W\rightarrow\infty)$, and depict the relation between $W_c$ and $V$ in Fig.~\ref{figetaw} (b). It is clear that $W_c\sim 1/V$ scales linearly as a function of $1/V$ similarly to $J_d$. From this respect, the well confined dynamics induced by disorder are related to the small second-order hopping, while a very small disorder is enough to localize quasiparticle NNDs with a small hopping amplitude $J_d$.

\subsection{Entanglement spectrum}
\begin{figure}[t!]
\centering
\includegraphics[width=0.48\textwidth]{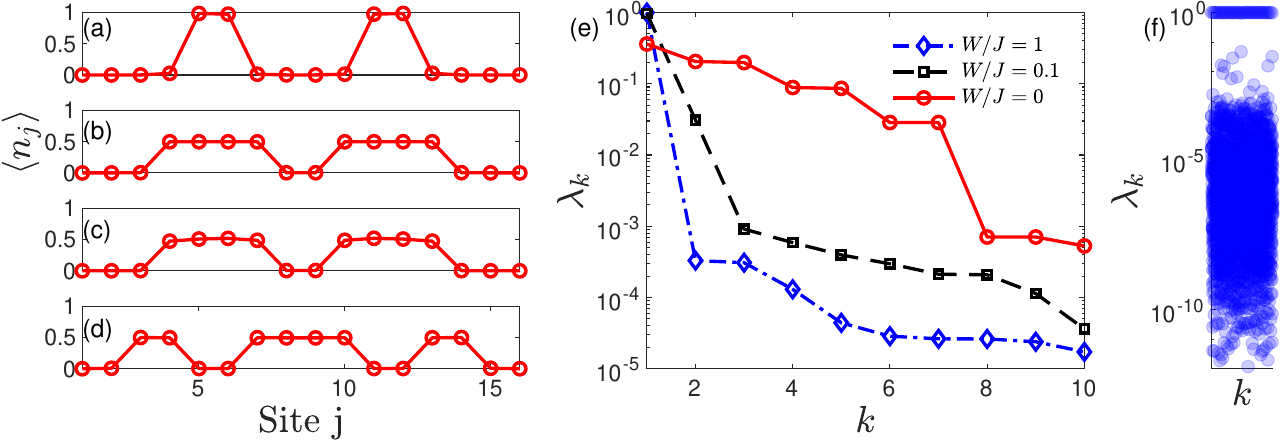}
\caption{(Color online) (a-d) Density distribution $\braket{n_j}$ of the largest four eigenstates of the reduced density matrix with $W=0$. (e) Ten largest eigenvalues $\lambda_k$ of the reduced density matrix in descending order for a single disorder realization with $W=0,0.1,1$, respectively. (f) Distribution of entanglement spectrum for disorder $W=1$ and $10^3$ disorder realizations. Reduced density matrices are calculated from long-time evolution $t=10^6$,  $V=40$, and the system size is $L=16$.}
\label{figes}
\end{figure}

The perfect dynamical confinement induced by the interplay between disorder and DDI can also be revealed from the viewpoint of the entanglement spectrum. In this section, we analyze the steady state $\ket{\psi(\infty)}$ of the system with $V=40$ after long-time evolution. Schmidt decomposition of the steady state reads $\ket{\psi(\infty)}=\sum_{ij}\delta_{ij}(\infty)\ket{\psi^A_i}\otimes\ket{\psi^B_j}$, from which we can define the reduced density matrix $\rho^A_{ii'}(\infty)=\sum_j \delta_{ij}(\infty) \delta^*_{i'j}(\infty)$. It can be diagonalized as $\rho^A=\lambda_k \ket{\psi_k^A}\bra{\psi_k^A}$ to obtain the entanglement spectrum $\lambda_k$. In numerical calculations, we use $t=10^6$ to approximate an infinite time. For $W=0$, the steady state is not dynamically confined and the first few values in the entanglement spectrum are about the same order of magnitude. We depict the first ten largest $\lambda_k$ in Fig.~\ref{figes} (e) with red cycles labeled line, and the corresponding density distributions of the four states with largest $\lambda_k$ are presented in Fig.~\ref{figes} (a-d), respectively. For a large DDI, although not confined due to the effective second-order coupling and center-of-mass motion, the state is kept in a paired NND basis where particles prefer to be in neighboring sites. While for a small disorder $W=0.1$, the entanglement spectrum begins to separate in Fig.~\ref{figes} (e) with black squares labeled curve, and the density distribution of the largest eigenstate is the same as in Fig.~\ref{figes} (a) which means the steady state is close to the initial state. For $W=1$ in Fig.~\ref{figes} (e) with blue diamond labeled line, the gap between the first and second largest $\lambda_k$ is even larger than that for $W=0.1$ and the dynamics are perfectly confined. To generalize the spectrum gap, we show the entanglement spectra of $10^3$ different disorder realizations for $W=1$ in Fig.~\ref{figes} (f). There is a distinct region between the largest $\lambda_k$ and the bulk spectrum, where only little points can be observed.

\subsection{Sample-to-sample fluctuations}
\begin{figure}[t!]
\centering
\includegraphics[width=0.48\textwidth]{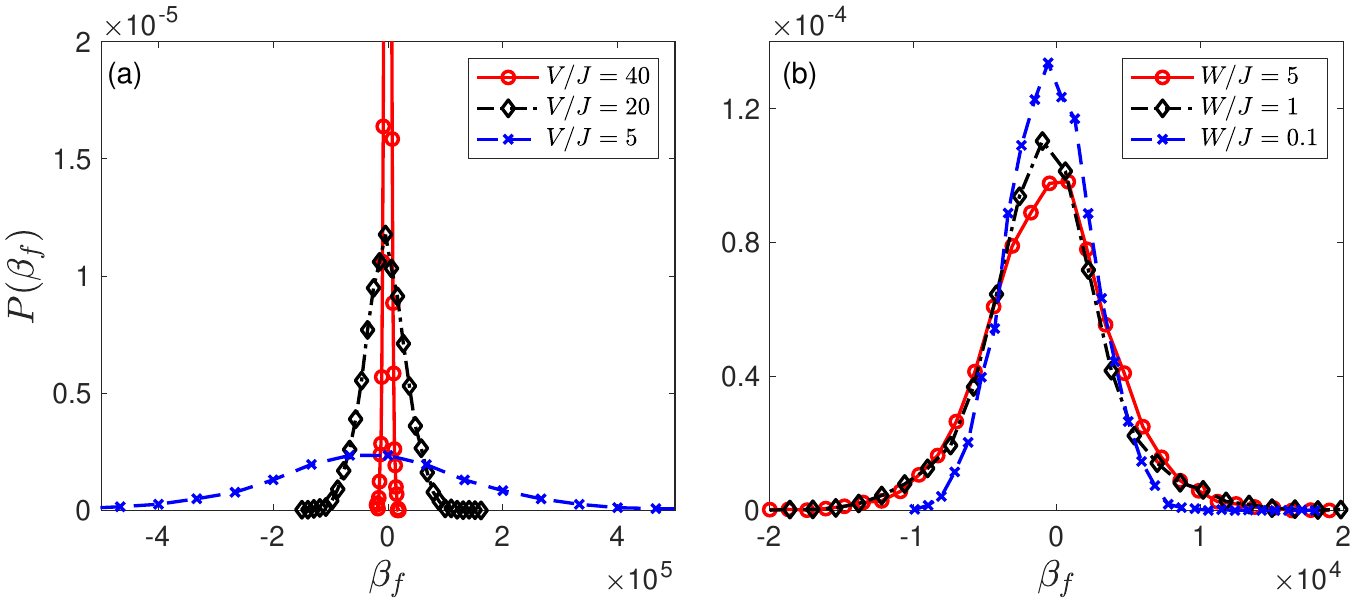}
\caption{(Color online) Probability distribution $P(\beta_f)$ of the effective exponent of $\eta$ in the long-time limit for (a) $W=5$ with $V=40$ (red cycles), $V=20$ (black diamonds), $V=5$ (blue crosses), and (b) $V=40$ with $W=5$ (red cycles), $W=1$ (black diamonds), $W=0.1$ (blue crosses). The system size is $L=16$, and $10^4$ disorder samples are used.}
\label{figssf}
\end{figure}

In the study of disorder induced localization, the average value of a physical quantity remains constant while the  distribution of that quantity is largely dominated by sample-to-sample fluctuations~\cite{PhysRevX.5.041047,PhysRevLett.117.160601,PhysRevB.94.045126,geraedts2017characterizing,PhysRevB.106.014201}. In the interplay of disorder and DDI, two types of localization mechanism may impact fluctuations differently. We further investigate the sample-to-sample fluctuation of the inhomogeneity parameter $\eta$. The effective exponent of $\eta$ in the long-time limit is given by
\begin{equation}
\beta_f=\frac{\partial \log_{10} \eta(t)}{\partial \log_{10} t}|_{t\rightarrow \infty},
\end{equation}
which characterize the dynamics of $\eta$ for an individual disorder configuration. In numerical calculations, $t=10^6$ is used to approach the long-time limit. We present the probability distribution $P(\beta_f)$ in Fig.~\ref{figssf} (a) for $W=5$ and various $V$'s, and in Fig~\ref{figssf} (b) for $V=40$ and several $W$'s. It is clearly illustrated by the blue cross labeled line in Fig~\ref{figssf} (a) that for strong disorder $W=5$ and small DDI $V=5$, the distribution is broad for the disorder dominated localization. Distribution of $P(\beta_f)$ shrinks when increasing $V$, and the DDI dominated confinements have much smaller sample-to-sample fluctuations. In Fig.~\ref{figssf} (b), we show the probability distribution $P(\beta_f)$ for $V=40$ dominated dynamical confinements with small $W=0.1$ to large $W=5$ disorders. The bandwidth does not increase significantly which reveals the stability of dynamical confinements with the existence of disorder. {\color{black}The sample-to-sample fluctuations between different disorder configurations for small $V$s are times larger than $V=40$ when $W=5$, which explains the quite large standard errors of mean values for those black, blue, and red curves shown in Figs.~\ref{figee} (c), \ref{figotoc} (c), \ref{figeta} (c), and \ref{figrp} (c).}

\section{\label{sec5}Quasiperiodic disorder and half-filling cases}
\begin{figure}[t!]
\centering
\includegraphics[width=0.48\textwidth]{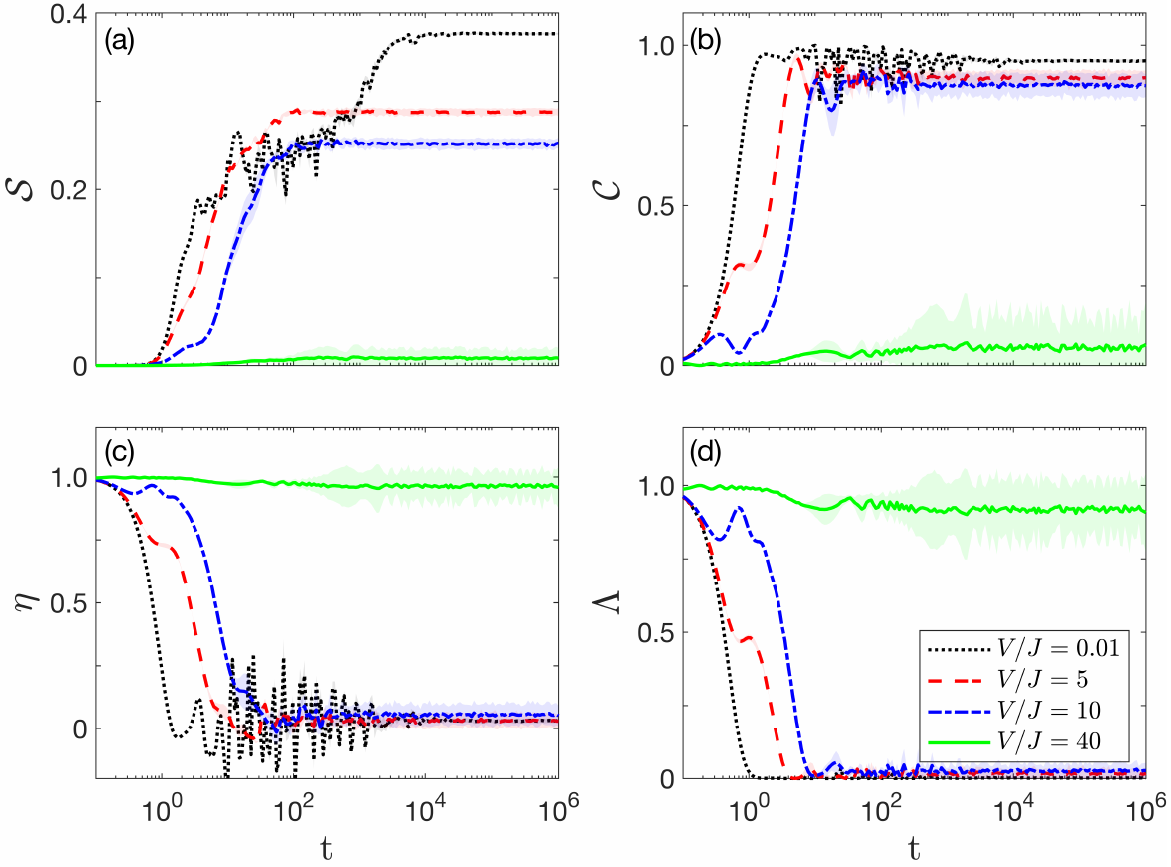}
\caption{(Color online) Long-time dynamics of (a) half-chain entanglement $\mathcal{S}$, (b) OTOC $\mathcal{C}$, (c) inhomogeneity parameter $\eta$, and (d) return probability $\Lambda$ for $V=0.01$ (black dotted curve), $V=5$ (red dashed curve), $V=10$ (blue dot-dashed curve), and $V=40$ (green solid curve). Faded color are the standard error between different offset phase $\varphi$. Data are averaged from 100 disorder realizations, and the system length is $L=20$ with $W=0.1$.}
\label{figqp}
\end{figure}
{\color{black}This section is devoted to extrapolating our results to systems with correlated disorders and more particles. We first show that the dynamically stabilized dimers can still be held with the interplay between quasiperiodic disorder effect and dipolar interaction. Then, we extend our two-dimer systems to the half-filling ones and reveal the stabilized quasiparticles in the presence of a very weak disorder.}

The quenched disorder uniformly distributed between $[-W, W]$ is an uncorrelated disorder, and in this section we further consider correlated disorders. The quasiperiodic disorder is a long-range correlated disorder that can also induce localization of dynamics. In Hamiltonian (\ref{ebhm}), the quasiperiodic on-site potential $\epsilon_i=W\cos(2\pi\alpha i+ \varphi)$ is used instead of the random $\epsilon_i$. $W$ is the quasiperiodic disorder strength, $\alpha=(\sqrt 5-1)/2$ is an irrational number close to the golden ratio, and $\varphi$ is an offset phase randomly chosen in range $[0, 2\pi)$ for sampling when the system's size is finite. Using the same parameters as in Figs.~\ref{figee}-\ref{figrp}, we plot the dynamics of half-chain entanglement, OTOC, inhomogeneity parameter, and return probability for the typical case where $W=0.1$ in Fig.\ref{figqp}. Those green solid lines indicate that the small quasiperiodic disorder can also stabilize the confined quasiparticle dynamics for large DDI $V=40$. While other curves reveal that the disorder induced localization is ignorable if $V$ is not strong enough. This typical behavior is similar to the random disorder, which extends our result from the uncorrelated disorder to the correlated disorder situation.

\begin{figure}[t!]
\centering
\includegraphics[width=0.48\textwidth]{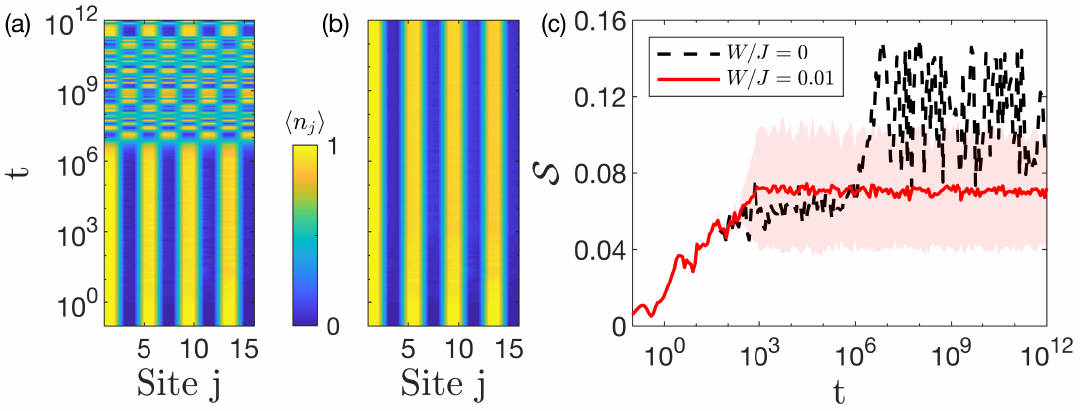}
\caption{\color{black}(Color online) Long-time dynamics of the density distribution $\braket{n_j}$ for half-filling systems without disorders (a), and with very weak disorders $W=0.01$ (b). (c) Evolution of half-chain entanglement $\mathcal{S}$ for $W=0$ (black dashed curve), and $W=0.01$ (red solid curve). Faded color are the standard error between different disorder configurations. Data are averaged from 100 disorder realizations, and the system length is $L=16$ with $V=20$.}
\label{fighf}
\end{figure}

{\color{black}Then we consider the half-filling case with the system size $L=16$ and the initial state containing $N_b=8$ bosons. The initial state has four dimers and is equally separated, which reads $\ket{\psi^h_0}=\ket{1100110011001100}$ in the Fock state basis. In this initial state, the distance between dimers is smaller compared with the previous configuration, where we use a smaller $V=20$ dipolar interaction strength. Despite being smaller, this interaction strength is sufficient to observe the confined dynamics. In Fig.~\ref{fighf}, we plot the long-time evolution of density distribution $\braket{n_j}$ and half-chain entanglement $\mathcal{S}$. For the clean system in Fig.~\ref{fighf} (a), the density distribution reveals the confined dimers for a long period up to $t\approx 10^7$ before the quasiparticles begin to move. In Fig.~\ref{fighf} (b), we can see that a very small disorder $W=0.01$ can significantly stabilize the confined dimers and the quasiparticles localized at their initial position for all the periods studied. Such a small disorder is not sufficient to induce localization if no dimer is present. Fig.~\ref{fighf} (c) shows the growth of half-chain entanglement for $W=0$ and $W=0.01$. For the clean system, the entanglement encounters a steep increasing period corresponding to the moving of quasiparticles at $t\approx 10^7$. For the stabilized confinement case ($W=0.01$), the entanglement saturates to a value smaller than that of the system with $W=0$. Our results suggest that the stabilizing mechanism induced by the interplay between disorder and dipolar interaction holds for systems with more particles.}

\section{\label{secf}Discussion and conclusion}
{\color{black}Before concluding, we emphasize the importance of investigating the interplay between dipolar interaction and disorder. Long-range interactions naturally occur in recent experimental quantum many-body systems such as cold gases, trapped ions, and engineered Rydberg atoms~\cite{kinoshita2006quantum,gring2012relaxation,richerme2014non,garttner2017measuring,bernien2017probing}.
On-site potentials with disorders or quasiperiodic disorders can also be implemented due to recent developments in MBL experiments~\cite{science.aao1401,PhysRevLett.122.170403}. Thus, the intriguing interplay between interaction- and disorder-induced quasiparticle localization is within the near-future quench dynamics in experiments. On the other hand, our findings provide a feasible technique in those long-range interacting systems to preserve the initially prepared quantum states for as long as possible. By introducing disorder on purpose, these systems can achieve longer decoherent time and realize more quantum tasks.
}

To summarize, we have explored the confined quasiparticle dynamics of the polar boson gas in a 1D lattice with both DDI and disorder. Several physical characteristics, such as the half-chain entanglement, the OTOC, the inhomogeneity parameter, and the return probability, have been numerically calculated. In the clean limit, four quantities evinced the thermalization of the system in the long-time limit. It was found that a small disorder, which is proportional to the inverse of DDI strength and comparable to the effective second-order hopping of dimers between nearest-neighbor sites, can stabilize the dynamical confinement. The entanglement spectra of perfect confined states indicated a significant gap between the first two largest eigenvalues. We have also unveiled that the DDI can reduce sample-to-sample fluctuations. Finally, the correlated quasiperiodic disorder {\color{black}and half-filling ystems have} been considered and similar findings have been established.

\begin{acknowledgments}
This work was supported by the NSFC (Grant No. 12104166),  and the Guangdong Basic and Applied Basic Research Foundation (Grants No. 2020A1515110290).
\end{acknowledgments}

{\color{black}
\appendix*
\section{Numerical method}
There are many approximation methods to calculate the time-dependent wave function $\ket{\psi(t)}=\mathrm{exp}(-iHt)\ket{\psi_0}$  such as the Krylov-subspace approach, Chebyshev polynomial approach, and tensor network base methods like the time-evolution block-decimation (TEBD) method and time-dependent variational principle (TDVP) method. Typically, these methods can evaluate time-dependent wave functions flexibly for a moderately long time from $t\approx 10^3$ to $t\approx 10^4$. However, we need to access physical properties of our systems with evolution times up to $t=10^{12}$, where the full exact diagonalization is the only appropriate method. The Hamiltonian matrix $H$ is represented under the Fock state basis with U(1) symmetry where the boson particle number is conserved. To calculate the time-dependent wave function, we first full diagnoalize $H=\sum_iE_i\ket{\psi_i}\bra{\psi_i}$ where $E_i$ and $\ket{\psi_i}$ are i-th eigenenergy and eigenfunction, respectively. The time-dependent wave function then reads $\ket{\psi(t)}=\sum_ie^{-iE_it}\braket{\psi_i|\psi_0}\ket{\psi_i}$. For any time $t$ and a disorder realization $d$, we calculate $\ket{\psi(t)}_d$ and its physical quantity $\braket{O(t)}_d$. All quantities are averaged over $100$ disorder realizations with their mean values plotted as lines and standard errors of the mean values plotted as the error bars.
}

 \bibliography{reference}

\end{document}